# Robust Optimization for Spot Scanning Proton Therapy based on Dose-Linear Energy Transfer (LET) Volume Constraints

**Authors:**

Jingyuan Chen, PhD[1], Yunze Yang, PhD[2], Hongying Feng, PhD[1,3,4], Lian Zhang, PhD[1], Carlos E. Vargas, MD[1], Nathan Y. Yu, MD[1], Jean-Claude M. Rwigema, MD[1], Sameer R. Keole, MD[1], Sujay A. Vora, MD[1], Jiajian Shen, PhD[1], Wei Liu, PhD[1]

[1]Department of Radiation Oncology, Mayo Clinic, Phoenix, AZ 85054, USA

[2]Department of Radiation Oncology, the University of Miami, FL 33136

[3]College of Mechanical and Power Engineering, China Three Gorges University, Yichang, Hubei 443002, People's Republic of China

[4]Department of Radiation Oncology, Guangzhou Concord Cancer Center, Guangzhou, Guangdong, 510555, People's Republic of China

**Abstract**

**Purpose:** Historically, spot scanning proton therapy (SSPT) treatment planning utilizes dose volume constraints and linear-energy-transfer (LET) volume constraints separately to balance tumor control and organs-at-risk (OARs) protection. We propose a novel dose-LET volume constraint (DLVC)-based robust optimization (DLVCRO) method for SSPT in treating prostate cancer to obtain a desirable joint dose and LET distribution to minimize adverse events (AEs).

**Methods**: DLVCRO treats DLVC as soft constraints that control the shapes of the dose-LET volume histogram (DLVH) curves. It minimizes the overlap of high LET and high dose in OARs and redistributes high LET from OARs to targets in a user defined way. Ten prostate cancer patients were included in this retrospective study. Rectum and bladder were considered as OARs. DLVCRO was compared with the conventional robust optimization (RO) method. Plan robustness was quantified using the worst-case analysis method. Besides the dose-volume histogram (DVH) indices, the analogous LET-volume histogram (LETVH) and extra-biological-dose (the product of per voxel dose and LET)-volume histogram (xBDVH) indices characterizing the joint dose/LET distributions were also used. The Wilcoxon signed rank test was performed to measure statistical significance.

**Results:** In the nominal scenario, DLVCRO significantly improved dose, LET and xBD distributions to protect OARs compared with RO (rectum: V70Gy: 3.07% vs. 2.90%, $p = .0063$, RO vs. DLVCRO; $\text{LET}_{max}$ $\left(\frac{\text{keV}}{\text{um}}\right)$: 11.53 vs. 9.44, $p = .0101$; $\text{xBD}_{max}$ (Gy $\cdot \frac{\text{keV}}{\text{um}}$): 420.55 vs. 398.79, $p = .0086$; bladder: V40Gy: 8.79 % vs. 8.46%, $p = .0287$; V65Gy: 4.82% vs. 4.61%, $p$

= .0032; $LET_{max}$ 8.97 vs. 7.51, $p$ = .0047, $xBD_{max}$ 490.11 vs. 476.71, $p$ = .0641). The physical dose distributions in targets are comparable ($D_{98\%}$: 98.57% vs. 98.39%, RO vs. DLVCRO; $p$ = .0805; CTV $D_{2\%} - D_{98\%}$: 7.10% vs. 7.75%; $p$ = .4624). In the worst-case scenario, DLVCRO markedly enhanced OAR protection (more robust) while maintaining almost the same plan robustness in target dose coverage and homogeneity.

**Conclusion**: DLVCRO upgrades 2D DVH-based to 3D DLVH-based treatment planning to adjust dose/LET distributions simultaneously and robustly. DLVCRO is potentially a powerful tool to improve patient outcomes in SSPT.

# Introduction:

Spot scanning proton therapy (SSPT) is one of the most advanced forms of proton therapy (PT)[1], utilizing pencil-beam scanning to precisely irradiate tumors in three dimensions[2-4]. This technique optimizes tumor control and the protection of adjacent organs at risk (OARs) by exploiting the unique physics properties of the Bragg peak and effectively modulating the intensities of beamlets[5-7]. Despite its dosimetric advantages, SSPT faces several challenges, such as plan robustness[8-21] and variable relative biological effectiveness (RBE)[22, 23].

Unlike photons, protons release most of their energy over a short distance, resulting in high linear energy transfer (LET)[24-26] near the distal end of the Bragg Peak. In clinical practice, a constant RBE of 1.1 is widely adopted as a coarse approximation to describe the RBE effect of protons compared with photons[27]. To study the relation between RBE and high LET, numerous in vitro[27-31] and in vivo[24, 25, 32, 33] studies have been conducted, and several phenomenological[34-37] and mechanism-related[38-43] RBE models have been proposed. Most of these phenomenological models were developed based on the linear-quadratic models, with the LET effect assumed as a function of cell radiation sensitivity change (α and β values) due to protons. The coefficients relative to photons are obtained by fitting the in vitro clonogenic cell survival data. Accordingly, besides the common dose volume constraints (DVCs) guided treatment planning strategies[44], various LET-guided or RBE-guided treatment planning strategies have been developed to improve clinical outcomes[8, 9, 23, 28, 34, 45-52].

Although the proton therapy community has well recognized the variation in RBE induced by high LET[27], the precise synergistic biological effects of dose and LET remain unclear[35]. The systematic comparisons revealed considerable variation among in vitro measurement results and, consequently, in the RBE estimations using different models[53]. Recent studies showed the synergistic biological effects of dose and LET in patient outcomes[54-57]. For example, the synergistic effects of dose and LET have been crucial in initiating AEs, as observed in studies concerning mandible osteoradionecrosis in head and neck cancer[56, 58, 59] and rib fractures in breast cancer patients receiving SSPT.[57, 60-62]

The methods to optimize dose and LET distributions separately through DVCs and LET volume constraints (LETVCs) reported in prior studies therefore cannot effectively address the synergistic biological effects between dose and LET in SSPT. To improve clinical outcomes and minimize AEs, the development of a treatment planning method that considers the synergistic effects of dose and LET in SSPT is crucial.

The newly introduced dose-LET volume histogram (DLVH) serves as an effective tool for integrating physical dose and LET to assess their synergistic biological effects in SSPT.[55, 56, 57, 63] This method improves upon traditional dose-volume histograms (DVHs) by including LET as an independent variable alongside the physical dose. DVHs typically represent three-dimensional dose distributions in a two-dimensional format, simplifying spatial data at the cost of losing certain spatial information of the dose distribution[64]. The DLVH, along with dose-LET volume constraints (DLVCs), were proven to be predictive of clinical outcomes, such as rectal bleeding in prostate cancer patients treated with SSPT[55].

SSPT is highly sensitive to range and setup uncertainties[18, 19]. These uncertainties stem from various sources, including daily patient alignment, the conversion of Hounsfield Units to stopping power, artifacts present in computed tomography scans, patient anatomic changes over time, and other factors[65-67]. Neglecting these uncertainties can lead to significant degradation of dose distributions for SSPT treatment plans[2068]. Additionally, LET distributions will also be altered in different uncertainty scenarios, which may result in unexpectedly high RBE doses to OARs [9, 28, 34, 45, 46, 69, 70]. Therefore, it is important to incorporate these uncertainties into the optimization of both physical and radiobiological doses, including LET distributions. Investigators have studied the impact of uncertainties upon dose and LET distributions and developed dose-related [8, 71] or LET-related robust optimization (RO) to improve the robustness of dose and LET distributions in the presence of uncertainties[9, 10, 72].

In this study, we introduced a novel robust optimization method based on DLVCs. This method significantly advances over the traditional DVC and LETVC-based robust optimization methods by employing constraints containing the synergistic effects of both dose and LET, rather than addressing them separately. We had applied this method to prostate cancer patients treated with SSPT as an example application since the DLVCs for rectal bleeding in prostate cancer patients treated with SSPT have been derived from the prior studies[55]. The proposed method has successfully minimized the overlap of high LET and high dose in rectum and bladder and redistribute high LET from rectum and bladder to targets with slight or little sacrifice of target physical dose and plan robustness.

## Methods and Materials

### Dose–LET volume histogram (DLVH) and extra biological dose (xBD)

DLVH has been recently introduced as a novel tool to facilitate analysis of the LET-enhancing effect for AE initialization in SSPT[55, 56, 57, 63]. The DLVH is constructed as a three-dimensional cumulative volume histogram that simultaneously displays the distributions of both dose and LET (dose-averaged) within a specified structure. Figure 1(c) illustrates a prototype of the three-dimensional DLVH structure. In this histogram, the axes represent the physical dose (measured in Gy) and the LET (measured in keV/μm), while the third axis corresponds to the normalized volume of the structure. The unit of the physical dose is Gy because we do not consider RBE here, while in other places we have used Gy[RBE=1.1] as the unit of dose, since this is commonly used in clinics. The DLVH index quantifies the volume $V_{D,LET}(d,l)$, defined as:

$$V_{D,LET}(d,l) = V(D \geq d, LET \geq l)$$

Here, $V(D \geq d, LET \geq l)$ denotes the cumulative normalized volume of the structure receiving a dose of at least $d$ (Gy) and LET of at least $l$ (keV/μm).

Analogous to the well-known DVH[64], the DLVH extends the concept by incorporating LET. It distinctly advances beyond the capabilities of the traditional DVH and LET-volume histogram (LETVH), which merely provide separate assessments of the dose and LET distributions without considering their potential interplay.

For effective visualization, a two-dimensional iso-volume contour can be derived from the three-dimensional DLVH by projecting its iso-volume curves onto the dose-LET plane as shown in Fig. 1(b). In this representation, the X-axis denotes the physical dose (measured in Gy), and the Y-axis represents the LET (in keV/μm). For the specific structure, the contour lines on this plane illustrate varying percentage volumes. Figure 1(b) displays contour lines for 5%, 20%, 50% and 80% iso-volumes, providing a clear and quantifiable depiction of how dose and LET interplay within the specific structure.

The product of dose and LET, known as extra biological dose (xBD), has been frequently utilized as a surrogate for describing the synergistic effects of high LET and dose in proton therapy plan optimization and evaluation[23, 74]. In addition, xBD has been identified as a good dose and LET descriptive feature for AE initialization in SSPT[56, 57]. Furthermore, xBD-volume constraints have been derived to predict mandible osteoradionecrosis for SSPT in the treatment of head and neck cancer[56, 57]. Besides the DVH and the LETVH, the xBD volume histogram (xBDVH) was also adopted in this work to evaluate the biological effect of the generated SSPT plans.

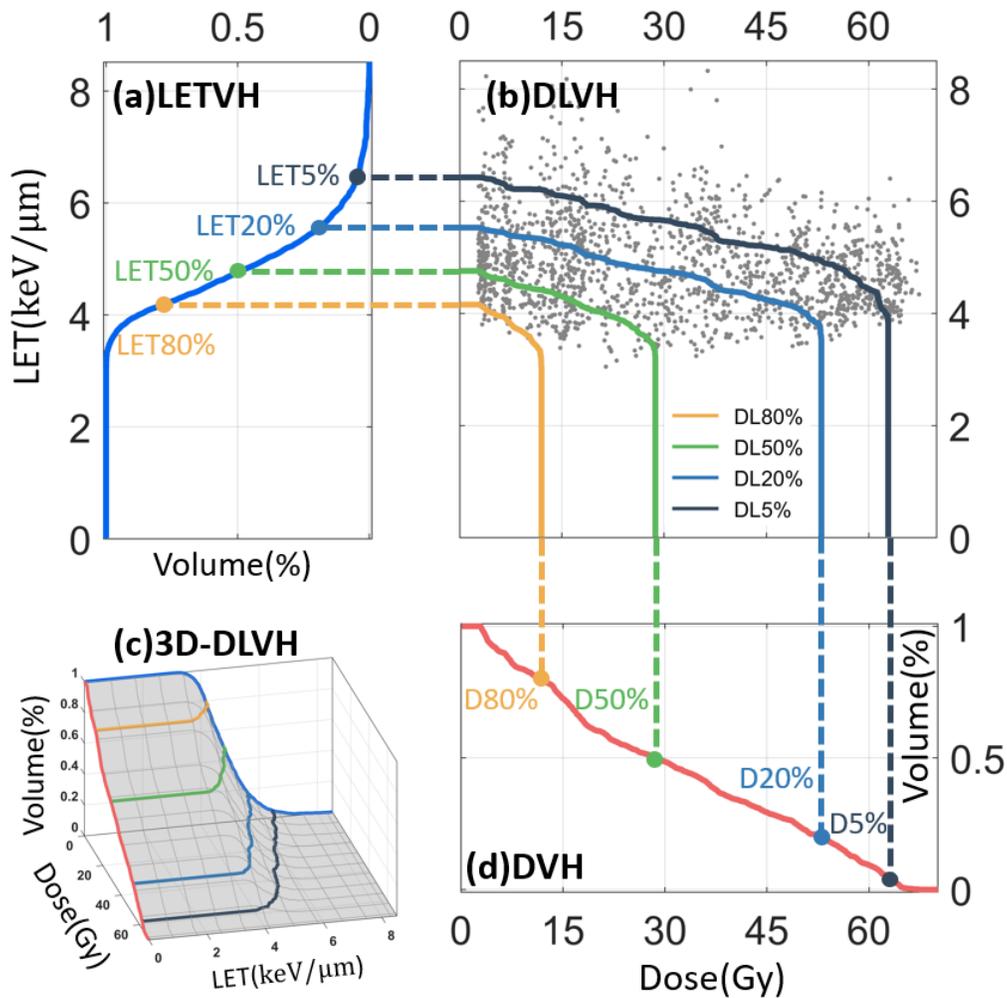

**Fig. 1.** Dose-linear-energy-transfer (LET) volume histogram (DLVH). An example of DLVH in rectum. Three-dimensional DLVH is sketched in (c). DL$v$% lines (solid lines with different colors) are the iso-volume contour of the 3D DLVH and represent the percentage volume of a structure that has a dose of at least $d$ Gy and an LET of at least $l$ keV/mm. The intersections of the 3D DLVH surface plot in the Dose-Volume plane is the Dose-volume histogram (DVH), as shown in (d). So as the LET-volume histogram (LETVH) in (a). The projected two-dimensional DLVH with the corresponding DL$v$% lines are shown in (b). The intersections of DL lines with the dose and LET axes are D$v$% in DVH and LET$v$% in LETVH, respectively. Each voxel of the structure was mapped to the 2D dose-LET plane indicated as gray dots in Figure (b).

**Dose-LET-Volume Constraints (DLVC)**

The current DVH or LETVH-based SSPT treatment planning methods to mitigate the impact of high LET[9, 28, 46, 47, 71, 75-80] are essentially of a 2D concept. Such methods depend on a two-way relationship of either between dose $d$ and volume $V$ or between LET $l$ and volume $V$, to change the dose or LET distribution separately, not simultaneously, to get a desirable dose/LET distribution (Fig. 2(a)). We will upgrade SSPT treatment planning from 2D to 3D with dose, LET, and volume considered via DLVH-based robust optimization to adjust the dose and LET distributions simultaneously (Fig. 2(b)).

The DLVCs will be implemented as "soft-constraints" in the objective function[44, 81, 82], which will modify the shape of the DLVH surface to satisfy DLVCs by only penalizing those voxels violating DLVCs, thus reducing the number of potential seed spots and leading to reduction of possible incidences of the corresponding AEs. In Fig. 2b, the yellow surface is the desired 3D DLVH surface, and the green surface is the current 3D DLVH surface. The red-pentagram indicates DLVC, i.e.,

$V_{desired}(f_1(d,l)) < V_1$ used in the optimization, where $f_1(d,l)$ is the curve **related to the specified DLVC, which could be derived from the DLVH-based patient outcomes study** and indicated by the black dash line. The DLVC limiting the overlap of high dose and high LET in OARs as shown in Fig. 2(b) is specified as: $V_{desired}(f_1(d,l)) < V_1$, *i.e.*, the relative volume receiving dose $\geq d$ and LET $\geq l$ correlated by the function of $f_1(d,l)$ (*black dash line*) should be no more than $V_1$.

We will follow the same concept of implementing DVCs based on DVHs to optimize physical dose distribution (Fig. 2(b)) to implement the DLVCs into the objective function[44]. For each iteration during the optimization, we will seek another curve $f_2(d,l)$ (*red dash line*), which is the intersection line of the current 3D DLVH surface and the parallel plane defined by $V = V_1$ (*purple plane*), so that in the current 3D DLVH surface: $V_{current}(f_2(d,l)) = V_1$. The two dash curves will then be projected (*vertical dash lines*) to the 2D dose-LET plane to form two solid curves. Only voxels with dose and LET values enclosed by these two curves and the axes of dose and LET, which forms a set of voxels, <u>VDLVC,</u> in the underlined term in Eq. (1) will be penalized for this structure in the optimization. If necessary, additional DLVCs can be introduced to further control the DLVH surface to achieve the desired dose and LET distributions simultaneously. Our model is thus able to redistribute dose and/or LET and minimize the possible incidence of AEs in OARs.

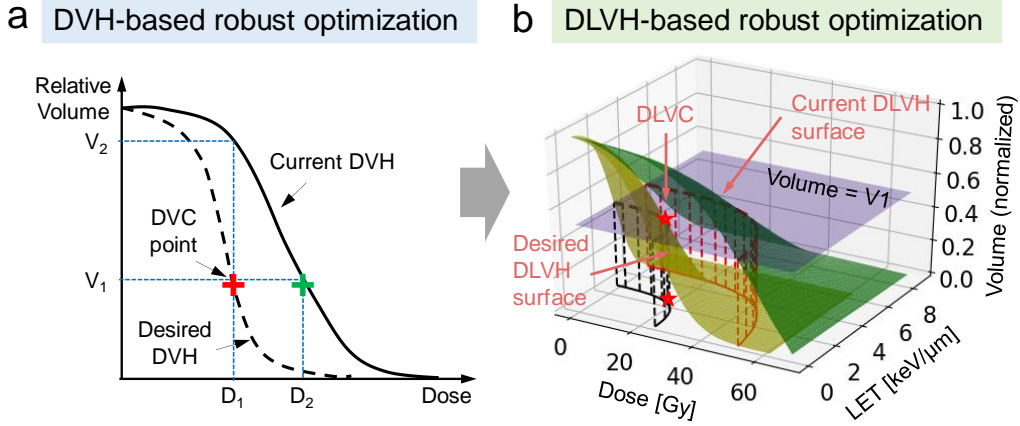

**Fig. 2.** DLVH-based optimization upgrades SSPT treatment planning from 2D to 3D. a) DVCs in conventional DVH-based optimization. Red cross indicates DVC, i.e., $V_{desired}(\geq D_1) \leq V_1$ used in the optimization. Only voxels with a dose between $D_1$ and $D_2$ are penalized in the DVC-based optimization. b) DLVCs in DLVH-based optimization. The yellow surface is the desired 3D DLVH surface, and the green surface is the current 3D DLVH surface. The red pentagram indicates DLVC, i.e., $V_{desired}(f_1(d,l)) \leq V_1$ used in the optimization, where $f_1(d,l)$ is the related to the specified DLVC, which could be derived from the DLVH-based patient outcomes study and is indicated by the black dash line. To implement the DLVC, for each iteration during the optimization, we seek another curve $f_2(d,l)$ indicated by the red dash line so that in the current DLVH surface $V_{current}(f_2(d,l)) = V_1$. The two dash curves will then be projected (vertical dash lines) to the dose-LET plane to form red and black solid curves. Only voxels with dose $d$ and LET $l$ values enclosed by these two curves and the axes of dose and LET will be penalized for this structure in the optimization. **Abbreviation:** DLVH = dose-linear-energy-transfer volume histogram; DVH = dose-volume histogram; SSPT = spot scanning proton therapy; DVC = dose-volume constraint; LET = linear-energy-transfer; DLVC = dose-linear-energy-transfer volume constraint.

## DLVCRO

The proposed DLVCRO is an enhancement of the voxel-wise worst-case RO method[8, 14-16, 72, 81, 83-87]. Nine scenarios and their combination were considered in the RO, including the nominal scenario, 2 scenarios due to range uncertainties, and 6 scenarios due to setup uncertainties. Range uncertainties were modeled by shifting the relative-to-water stopping power ratio conversion curve by ±3%, simulating maximum and minimum range uncertainties [20]. Setup uncertainties involve rigid shifts of the isocenter by 5 mm in the anteroposterior, superoinferior, and left-right Cartesian directions.

An additional "biological surrogate (BS)" term[8, 23] (underlined term in Eq. (1)) was introduced as indirect RBE optimization. The BS term, $BS_i^m = D_i^m + \sum_j c L_{i,j}^m IM_{i,j}^m \omega_j^2$, where the LET influence matrix[8, 9] $L_{i,j}^m$ describes the LET contribution of beamlet $j$ of unit intensity to voxel $i$ in uncertainty scenario $m$. $D_i^m$ is the dose at ith voxel for mth uncertainty scenario[88, 89] and can be calculated as: $D_i^m = \sum_j IM_{i,j}^m \omega_j^2$, where $\omega_j^2$ represents non-negative intensity weight of beamlet $j$. The dose influence matrix $IM_{i,j}^m$ describes the dose contribution of beamlet $j$ of unit intensity to voxel $i$ in uncertainty scenario $m$[15, 90]. $c$ is a scaling constant and is set to be 0.04 μm/keV[8, 23], and the second term can be regarded as extra biological dose (xBD) compared to the physical dose $D_i^m$.

A standard quadratic objective function of our proposed DLVCRO is designed as follows:

$$F_{obj}^{LET} = \sum_{i \in CTV} (p_{CTV}^{max}(D_i^{max} - D_{0,CTV}^{max})_+^2 + p_{CTV}^{min}(D_{0,CTV}^{min} - D_i^{min})_+^2)$$

$$+ \underline{\sum_{i \in VDLVC} p_{OAR}^{BS}(BS_i^{max} - BS_{0,OAR})_+^2} \quad (1)$$

The first 2 terms in Eq. (1) ensure that minimum and maximum physical dose constraints for the clinical target volume (CTV) are met. The Heavyside function $(x - x_0)_+$ returns $(x - x_0)$ if $x > x_0$, otherwise zero. Terms with subscript "0" ($D_{0,CTV}$, $D_{0,CTV}$, and $BS_{0,OAR}$) are the prescribed dose or BS. $p_{CTV}^{min}$, $p_{CTV}^{max}$, and $p_{OAR}^{BS}$ are penalty weights for each corresponding term. $D_i^{min}$, $D_i^{max}$, and $BS_i^{max}$ stand for the voxel-wise minimum or maximum value of the dose or BS among all the values corresponding to different uncertainty scenarios (voxel-wise worst-case dose/BS)[91].

The influence matrix $IM_{i,j}^m$, was calculated using an in-house semi-analytical dose engine that employs a modified ray-casting pencil beam algorithm with 3 lateral Gaussian components[92]. The LET influence matrix $L_{i,j}^m$ was calculated using an in-house developed LET calculation code[93]. This study utilized a dose-averaged LET, calculated with restricted stopping powers. The methodology for calculating LET parallels the pencil beam algorithm used in dose calculations. A three-dimensional LET calculation kernel was created to replace the standard dose calculation kernel, using the precomputed LET data from Monte Carlo simulations[93-96].

Optimization was performed using the limited-memory Broyden-Fletcher-Goldfarb-Shanno (LBFGS) algorithm, which was enhanced through parallel computation via a message passing interface[97]. In the objective function, the physical DVCs for the first two terms of Eq. (1) were treated as "soft constraints"[44, 81, 87]. Due to the nonconvex nature of the objective function, a "trial-and-error" method was employed to adjust the initial optimization parameters until a clinically acceptable plan was achieved.

**Patient selection and machine configuration**

We generated two SSPT plans using both the RO and DLVCRO methods for each of 10 prostate cancer patients. Each patient underwent treatment with plans featuring anti-parallel left and right fields, employing multi-field optimization in both the RO and DLVCRO approaches. All plans began optimization with identical initial conditions regarding beamlet intensities, number, energy, and positions. Subsequently, only beamlet intensities were optimized using the two methods under study.

Following optimization, a postprocessing procedure was applied to both the RO and LETRO plans to ensure deliverability by adhering to the minimum monitor unit (MU) limits. Specifically, any spot planned to deliver MUs below the minimum threshold (0.003 MU for our proton therapy machine) was adjusted upwards or downwards, depending on if it was above or below half the minimum MU limit. The dose grid voxel size was maintained at 2.5 × 2.5 × 2.5 mm throughout the study.

**Plan evaluation**

We evaluated CTV $D_{98\%}$—the minimum dose normalized by the prescription dose to cover 98% of the target—and $D_{2\%} - D_{98\%}$ for evaluating target dose coverage and homogeneity, respectively. To evaluate the OAR sparing, V40Gy[RBE=1.1] and V65Gy[RBE=1.1] of bladder and V70Gy[RBE=1.1] of rectum were measured. To ensure fairness in comparisons, CTV $D_{98\%}$ in the nominal scenario of each plan was normalized to the prescription dose, regardless of the optimization method used. LETVHs were utilized to compare LET distributions across different plans. Indices similar to those in DVH were employed to evaluate these distributions. We calculated CTV $LET_{98\%}$ and $LET_{mean}$ to determine target LET coverage and average LET, respectively, where $LET_{98\%}$ is defined as the minimum LET required to cover 98% of the target.

For LET protection in OARs, $LET_{max}$ for the bladder and rectum. $xBD_{mean}$ is calculated for CTV to evaluate the biological effects. The $xBD_{max}$ of bladder and rectum is calculated to indicating the biological effects.

**Plan robustness**

In assessing plan robustness, we considered setup uncertainties of 5 mm and range uncertainties of 3%, which are currently the standards for patients with prostate cancer undergoing radiation therapy with image guidance at our institution. A total of 21 uncertainty scenarios were analyzed, including nominal, minimum (-3%), and maximum (+3%) proton ranges, along with patient positions shifted in anteroposterior, superoinferior, and left-right directions (7 scenarios per proton range). We employed the worst-case scenario analysis method [98, 99]—recognized for its speed and efficiency—to evaluate plan robustness across these scenarios. The minimum CTV $D_{98\%}$ was used to assess the poorest target dose coverage, while the maximum CTV $D_{2\%} - D_{98\%}$ evaluated the least homogeneity in target dose. Furthermore, the V40Gy[RBE=1.1] and V65Gy[RBE=1.1] of bladder and V75Gy[RBE=1.1] of rectum were measured to determine OAR protection in the worst-case scenario. Corresponding LETVH and xBDVH indices were calculated to assess the robustness of the LET distribution and xBD distribution in the face of these uncertainties.

**Statistical analysis**

The Wilcoxon signed-rank test was utilized to compare all evaluation metrics, with calculations performed using SciPy Python library. *p*-values below 0.05 were deemed statistically significant.

In the box-and-whisker plot, data points falling outside 1.5 times the interquartile range above the upper quartile or below the lower quartile were identified as outliers.

## Results

**Physical dose plan quality**

We assessed plan quality in the nominal scenario between the two methods: RO and DLVCRO. As depicted in Fig.3, both methods produced SSPT plans with comparable target dose homogeneity (CTV $D_{2\%} - D_{98\%}$: 4.53% vs. 4.25%, RO vs. DLVCRO; $p$ = .3887). Furthermore, both RO and DLVCRO provided better OAR protection in the bladder and rectum, as reflected by the DVH indices (bladder V40Gy[RBE=1.1]: 8.79 % vs. 8.46%, $p$ = .0287; bladder V40Gy[RBE=1.1]: 4.82% vs. 4.61%, $p$ = .0032; rectum V40Gy[RBE=1.1]: 3.07% vs. 2.90%, $p$ = .0063, RO vs. DLVCRO). DLVCRO seems to generate SSPT plans with the similar plan quality in target as RO and better dose distribution to protect OARs than RO.

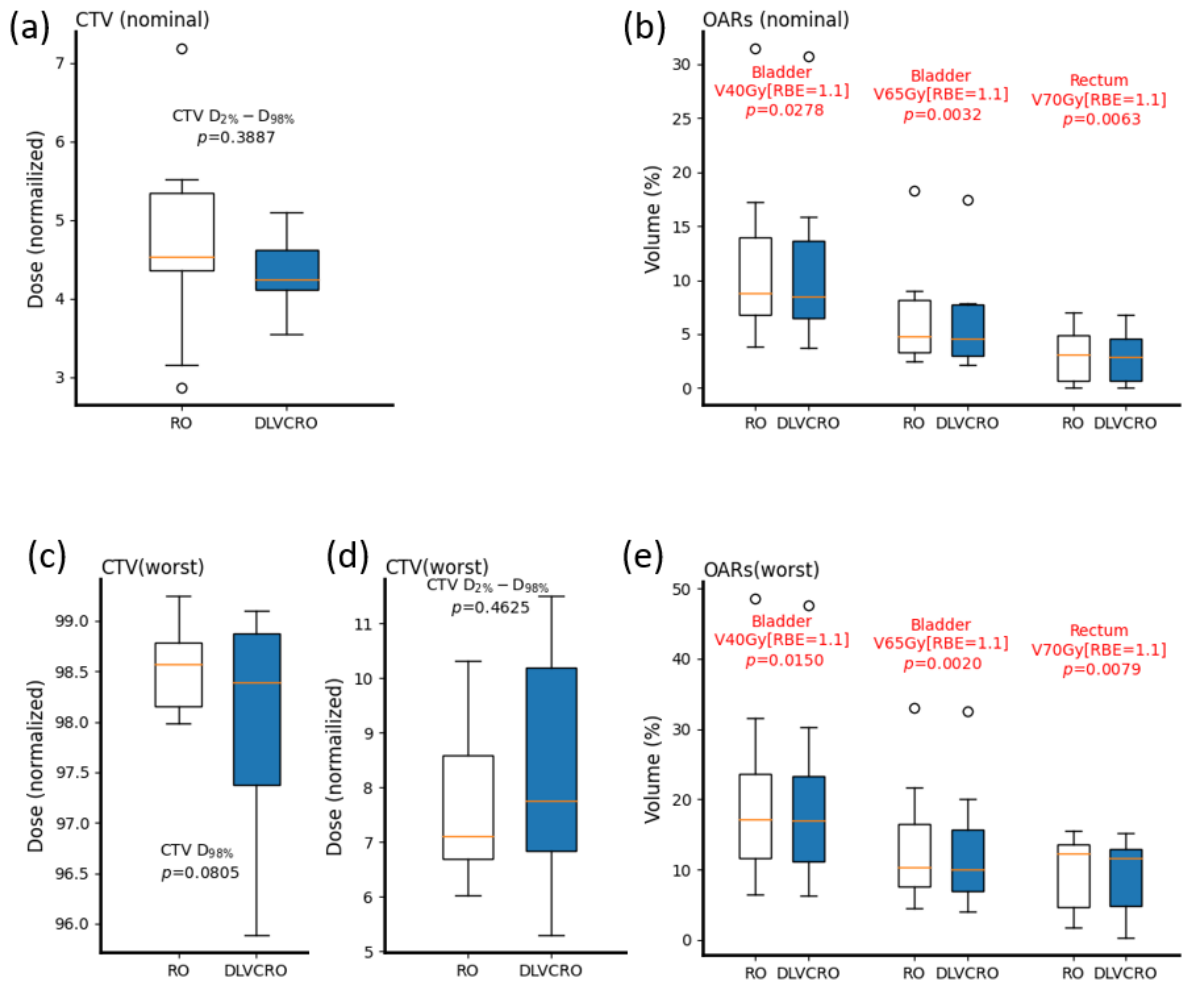

**Fig. 3.** Boxplots comparing dose-volume histogram indices of the treatment plans generated by RO and DLVCRO in the nominal scenarios and worst-case scenarios for all 10 patients (a) Normalized CTV $D_{2\%} - D_{98\%}$ in the nominal scenario. (b) Bladder V40Gy[RBE=1.1], V65Gy[RBE=1.1] and rectum V70Gy[RBE=1.1] in the nominal scenario. (c) Normalized CTV $D_{98\%}$ in the worst-case scenario. (d) Normalized CTV $D_{2\%} - D_{98\%}$ in the worst-case scenario. (e) Bladder V40Gy[RBE=1.1], V65Gy[RBE=1.1] and rectum V70Gy[RBE=1.1] in the worst case scenario. $p$-values less than 0.05 are highlighted, which are calculated from the Wilcoxon signed rank test. **Abbreviation:** RO = robust optimization; DLVCRO = dose-linear-energy-transfer volume constraint robust optimization; CTV = clinical target volume.

**Physical dose plan robustness**

With range and setup uncertainties accounted for, we calculated the DVH indices for the CTV and OARs in the worst-case scenario, as shown in Fig. 3. When compared to the RO method, plans developed using DLVCRO demonstrated similar robustness in target dose coverage and homogeneity, and improved OAR protection. (CTV $D_{98\%}$: 98.57% vs. 98.39%, RO vs. DLVCRO; $p = .0805$; CTV $D_{2\%} - D_{98\%}$: 7.10% vs. 7.75%; $p = .4624$; bladder V40Gy[RBE=1.1]: 17.21% vs. 16.93%, $p = .0150$; bladder V65Gy[RBE=1.1]: 10.42% vs. 9.98%, $p = .0020$; rectum V70Gy[RBE=1.1]: 12.24% vs. 11.68%, $p = .0079$). DLVCRO seems to generate more robust plans for OARs with similar plan robustness for targets compared to RO.

**LET distribution**

Figure 4 illustrates the LETVH indices of the plans generated using both methods under the nominal and worst-case scenarios. In nominal scenarios, DLVCRO provided significantly better LET protection for OARs and similar target LET coverage (unit: keV/μm; CTV $LET_{mean}$ 5.99 vs. 6.00, $p = .6566$ RO vs. DLVCRO; CTV $LET_{98\%}$ 4.65 vs. 4.65, $p = .3152$; bladder $LET_{max}$ 10.86 vs. 10.47, $p = .0040$, rectum $LET_{max}$ 8.97 vs. 7.51, $p = .0047$). In the worst-case scenarios, DLVCRO again showed improved LET protection for OARs and comparable target LET coverage (unit: keV/μm; CTV $LET_{mean}$ 5.28 vs. 5.31, $p = .3538$ RO vs. DLVCRO; CTV $LET_{98\%}$ 4.28 vs. 4.35, $p = .3036$; bladder $LET_{max}$ 12.00 vs. 11.39, $p = .0078$, rectum $LET_{max}$ 11.53 vs. 9.44, $p = .0101$). Compared to RO, DLVCRO seems to generate SSPT plans with better and more robust LET distribution for OARs and similar LET distribution for targets.

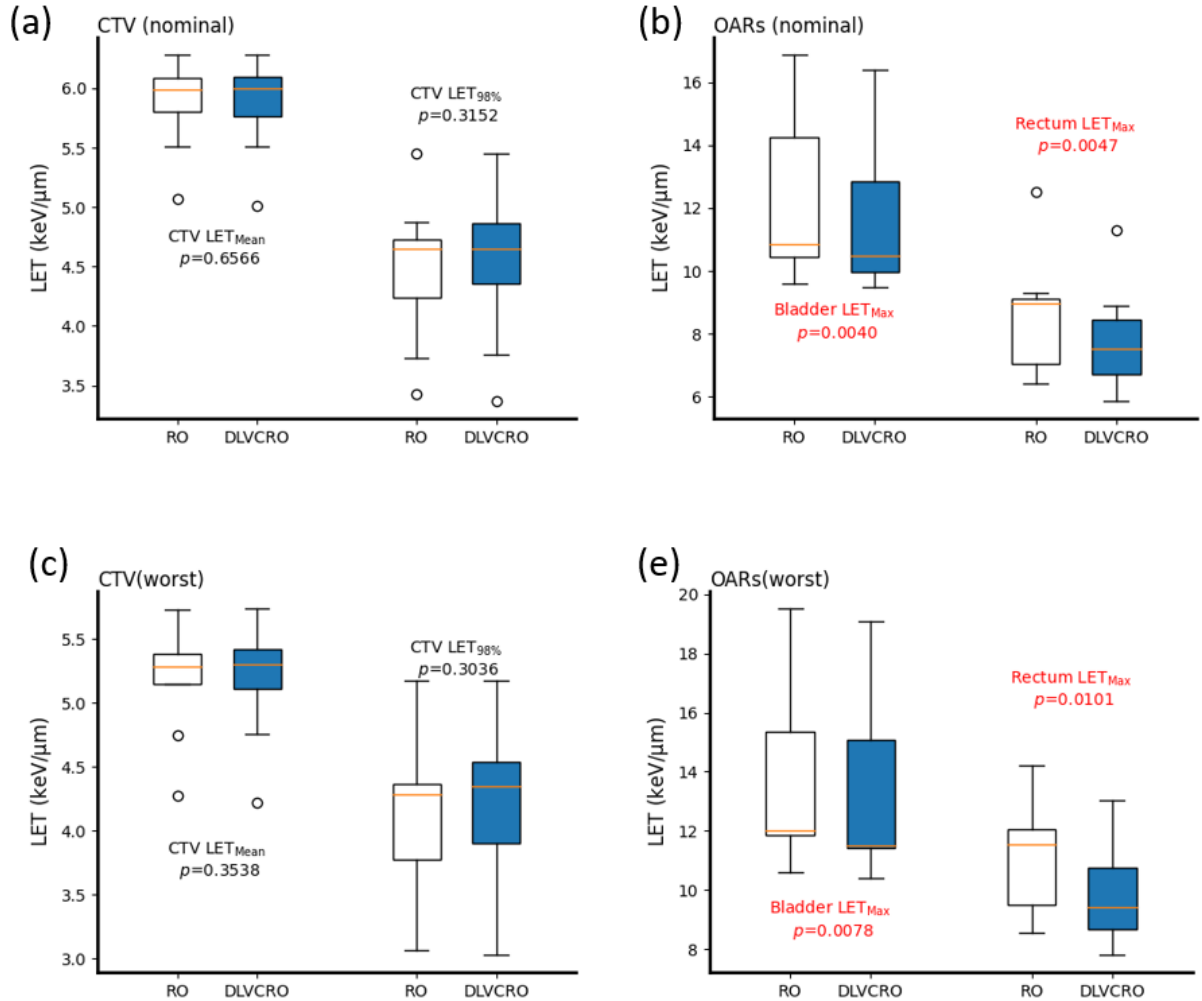

**Fig. 4.** Boxplots comparing LET-volume histogram indices of the treatment plans generated by RO and DLVCRO in the nominal scenarios and worst-case scenarios for all 10 patients (a) CTV $LET_{mean}$ and $LET_{98\%}$ in the nominal scenario. (b) Bladder $LET_{max}$ and rectum $LET_{max}$ in the nominal scenario. (c) CTV $LET_{mean}$ and $LET_{98\%}$ in the worst-case scenario. (d) Bladder $LET_{max}$ and rectum $LET_{max}$ in the worst-case scenario. $p$-values less than 0.05 are highlighted, which are calculated from Wilcoxon signed rank test. **Abbreviation:** LET = linear energy transfer; RO = robust optimization; DLVCRO = dose-linear-energy-transfer volume constraint robust optimization; CTV = clinical target volume.

**xBD distribution**

Figure 5 presents the xBDVH indices for plans generated by the two competing methods under both nominal and worst-case scenarios. Under nominal conditions, DLVCRO achieved superior xBD protection for OARs and equivalent target xBD coverage (unit: $Gy \cdot \frac{keV}{\mu m}$; CTV $xBD_{mean}$ 272.80 vs. 272.36, $p = .9552$ RO vs. DLVCRO; bladder $xBD_{max}$ 490.11 vs. 476.71, $p = .0641$, rectum $xBD_{max}$ 420.55 vs. 398.79, $p = .0086$) compared to RO. In the worst-case scenarios, DLVCRO continued to provide better OAR protection and similar target xBD coverage (unit: $Gy \cdot \frac{keV}{\mu m}$; CTV $xBD_{mean}$ 237.42 vs. 237.02, $p = .3902$ RO vs. DLVCRO; bladder $xBD_{max}$ 769.32 vs. 655.05, $p = .0176$, rectum $xBd_{max}$ 664.29 vs. 556.82, $p = .0052$). Compared to RO, DLVCRO seems to generate SSPT plans with better and more robust xBD distribution for OARs and similar xBD distribution for targets.

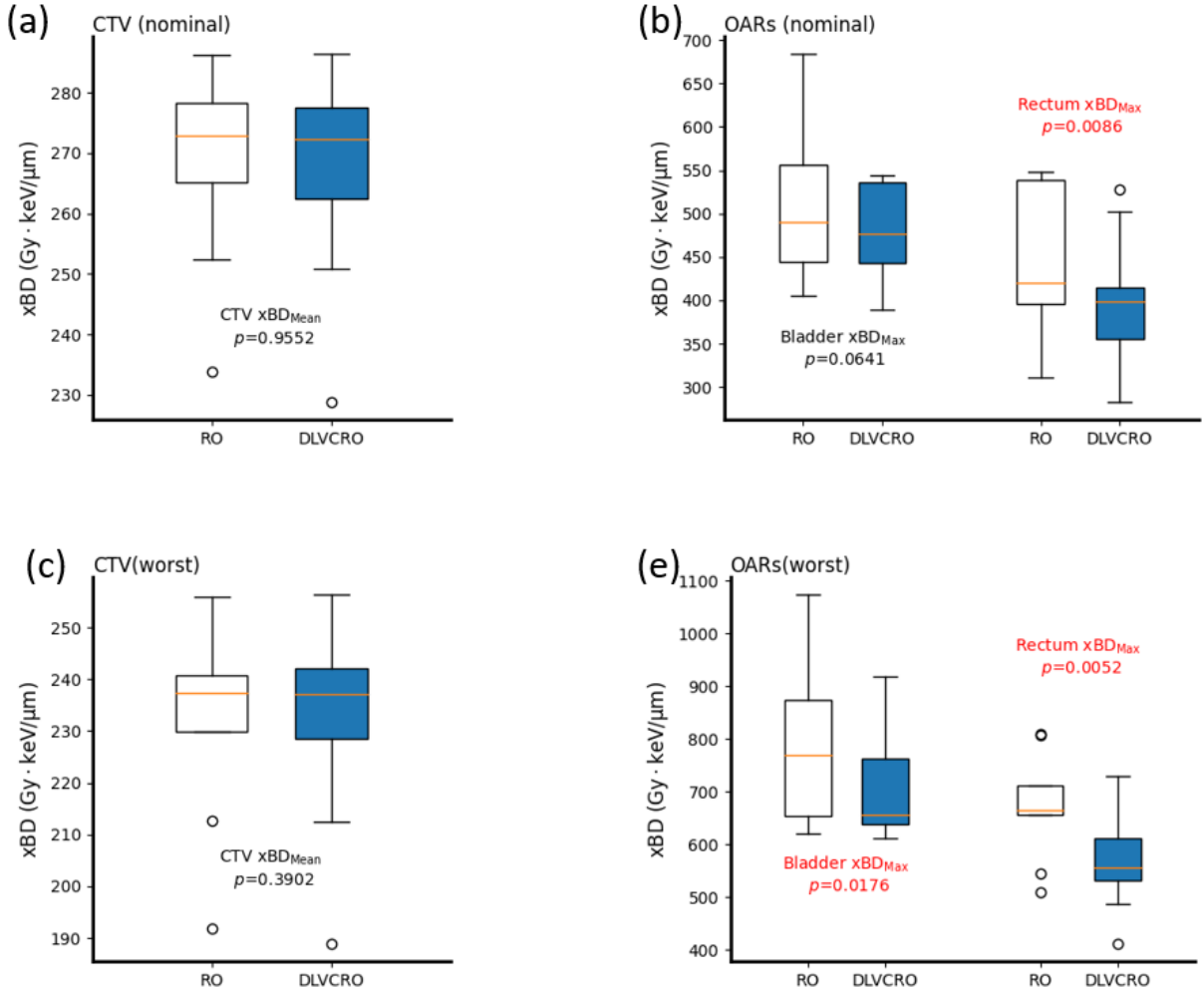

**Fig. 5.** Boxplots comparing xBD-volume histogram indices of the treatment plans generated by RO and DLVCRO in the nominal scenarios and worst-case scenarios for all 10 patients (a) CTV $xBD_{mean}$ in the nominal scenario. (b) Bladder $xBD_{max}$ and rectum $xBD_{max}$ in the nominal scenario. (c) CTV $xBD_{mean}$ in the worst-case scenario. (d) Bladder $xBD_{max}$ and rectum $xBD_{max}$ in the worst-case scenario. *p*-values less than 0.05 are highlighted, which are calculated from Wilcoxon signed rank test. **Abbreviation:** xBD = extra biological dose; RO = robust optimization; DLVCRO = dose-linear-energy-transfer volume constraint robust optimization; CTV = clinical target volume.

## DLVH, DVH and LETVH comparation

Figure 6 presents the comparison of the DLVHs for a representative patient, contrasting the RO (left) and the DLVCRO (right) methods. In DLVCRO, DLVCs for $V_{desired}(f_i(d, l)) \leq V_i$, at $V_i = 0$ are imposed on the corresponding OARs. These DLVCs are shown as the red dashed lines in both the RO and DLVCRO DLVHs for a clearer comparison. DLVCRO offers enhanced OAR protection by more effectively managing the overlap of high LET and high dose distributions within the OARs, as indicated by the yellow circles in Fig. 6(a) - (d). The DVH and LETVH for this patient are also provided in Fig. 6(e) and (f). DLVCRO results in lower LET exposure in OARs and increased LET within the CTV, while the physical doses in the CTV are analogous to those obtained with RO.

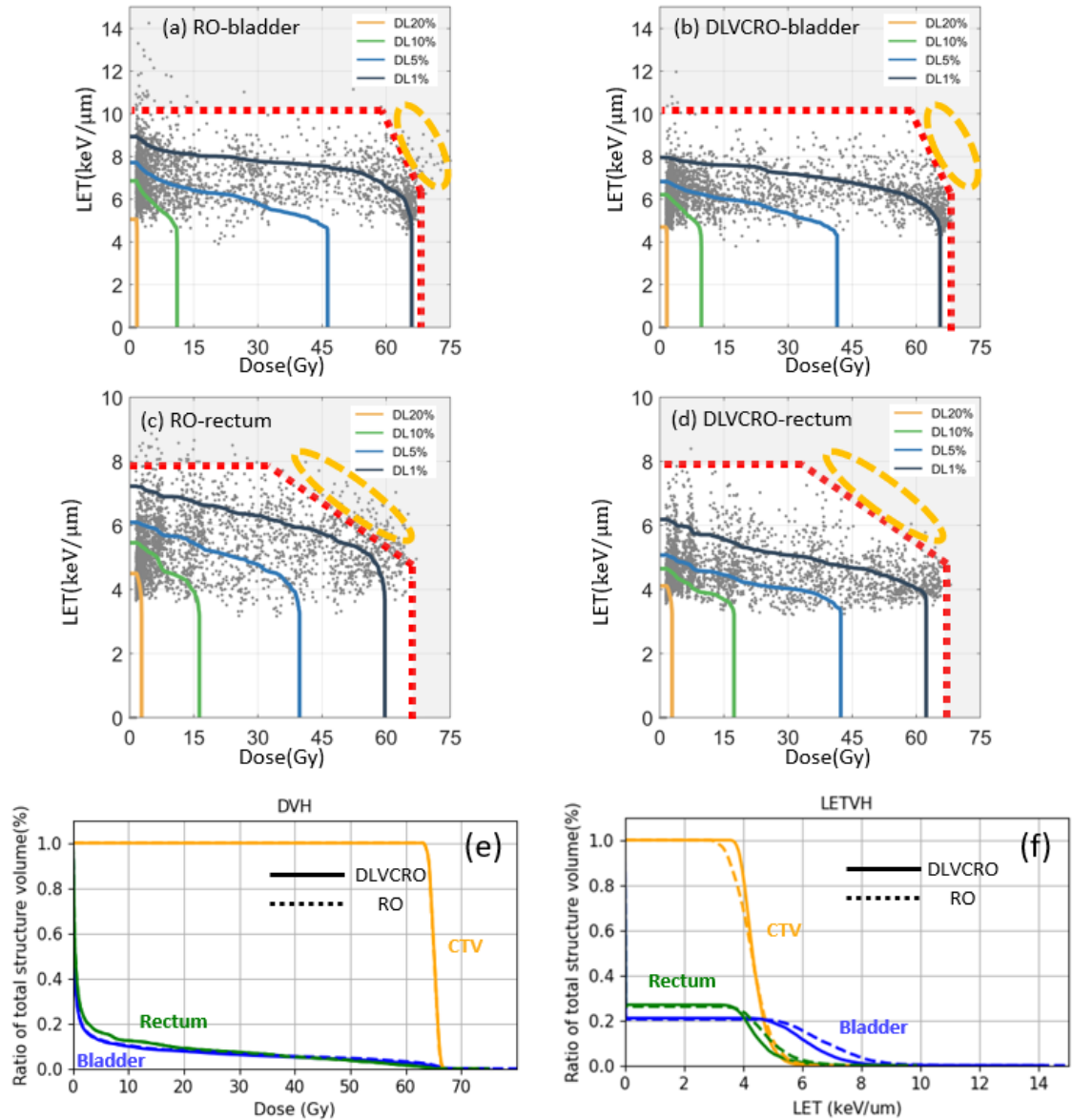

**Fig. 6.** Comparison of the LET and Dose in the DLVH, DVH and LETVH of the treatment plan generated by RO and DLVCRO for a representative patient. DLVHs of the bladder are generated by (a) RO and (b) DLVCRO, respectively. DLVHs of the rectum are generated by (c) RO and (d) DLVCRO, respectively. In the DLVHs of bladder and rectum, red dash lines of $V(f(d.l)) = 0\%$ indicate the DLVCs adopted in DLVCRO for bladder and rectum, respectively. Compared with RO, DLVCRO efficiently reduced the high LET and high dose regions, which are indicated by the yellow circles. In the DVH and LETVH (e, f), yellow is clinical target volume, blue is bladder and green is rectum; the solid line is DLVCRO, and the dashed line is RO.

**Abbreviations:** LET = linear energy transfer; DVH = dose-volume histogram; LETVH = linear energy transfer volume histogram; DLVH = dose-LET volume histogram; RO = robust optimization DLVCRO = dose-LET volume constraint robust optimization.

## Discussion

In this study, we introduced a novel optimization algorithm, DLVCRO, designed to robustly optimize both physical dose and LET distributions simultaneously and enhance OAR protection by implementing joint constraints of dose and LET. We employed DLVCRO to develop SSPT treatment plans for 10 patients with prostate cancer. The most significant finding is that DLVCRO substantially improves both dose and LET protection for OARs, albeit with a slight compromise in target physical dose distribution and robustness. Compared to the traditional robust optimization methods, our approach adopted DLVCs, which constrain the joint distribution of dose and LET. All DLVCs can be customized according to user specifications. By redistributing high LET from OARs to targets, our approach could potentially reduce the possible incidence rates of the corresponding AEs.

Unlike many current methods that integrate various RBE models in proton therapy treatment planning[27, 100, 101 8, 23, 28], our proposed DLVCRO method directly considers the synergistic effects of LET and dose without relying on any RBE models or assumptions, while an LET guided optimization method to indirectly account for the high RBE induced by high LET. Our method is also different from the prior LET-guided optimization methods to optimize the dose and LET distributions separately by simultaneously optimizing dose and LET distribution via a novel concept of DLVC. This approach may gain broader acceptance in the clinical community due to its alignment with prevailing dose-reporting practices[102, 103]. Moreover, our algorithm, which utilizes conventional nonlinear programming with a quadratic cost function, can be seamlessly integrated into the existing commercial treatment planning systems such as Eclipse (Varian Medical Systems, Palo Alto, CA), leveraging the same mathematical framework.

Compared to RO, DLVCRO achieved comparable plan quality in the nominal scenario with respect to target coverage and homogeneity (Fig. 3). And it has achieved comparable plan robustness for targets compared to RO (Fig. 3). Furthermore, our innovative DLVCRO method enhanced the protection of bladder and rectum in both the nominal and worst-case scenarios with respect to physical dose (Fig. 3), LET (Fig. 4), and xBD (Fig. 5). Compared to the conventional DVC and LETVC-guided methods to constrain the physical dose and LET distributions separately, our approach more effectively minimize the overlap of the high dose and high LET in OARs and redistribute high LET to targets. The DLVCs significantly reduce the synergistic effects of high dose and high LET in OARs to minimize the possible biological damaging to OARs while minimally impacting target physical dose and robustness. Unlike the conventional DVC and LETVC-guided methods that often involve irrelevant voxels with lower-dose and high-LET, as shown in Fig. 6, our approach achieves similar outcomes more effectively.

The solution space for SSPT optimizations is highly degenerate. Many plans maintain similar physical dose distributions and robustness, yet they differ in LET distribution—some have the desirable high LET in targets and low LET in OARs, while others have the opposite. Our innovative concept of DLVC-based robust optimization method, based on a novel tool of DLVH, steers the optimizer to generate SSPT plans with the optimal joint dose-LET distribution in a user-defined manner. Physical dose quantifies the energy deposited per mass unit, whereas LET measures the average energy loss per distance traveled by numerous particles[104]. This distinction allows for various joint dose-LET distributions without significantly altering the physical dose distribution.

The 10 patients with prostate cancer were treated with plans utilizing anti-parallel left and right fields. Multi-field optimization was implemented in both the RO and DLVCRO approaches. Due to the antiparallel treatment fields, which usually produce quite overall low LET distributions, and the anatomical geometrical relationship between targets and OARs in prostate cancer, it is always challenging to further evaluate LET within the targets through optimization. Additionally, both DLVCRO and RO could generate highly modulated fields, which may heighten the susceptibility to the field misalignment. Therefore, it is important to consider the impact of the uncertainties through robust optimization in both cases.

This study has several limitations. Notably, the optimization framework did not consider beam angle selection, potentially limiting the optimality of the resulting SSPT plans with respect to both dose and LET distributions. Furthermore, calculations were based on an analytical dose and LET calculation engine, which is acceptable in prostate cancer with overall homogeneous anatomies. In subsequent research, our aim is to include beam angle selection in the DLVCRO framework and to apply a Monte Carlo (MC) dose and LET calculation engine[94-96] to improve the dose and LET calculation accuracy in more inhomogeneous disease sites such as head and neck cancer.

## Conclusions

In this study, DLVCRO was distinguished from the traditional RO by optimizing the joint dose and LET distributions simultaneously to account for the synergistic effects of dose and LET in prostate cancer patients treated with SSPT. DLVCRO demonstrated superior performance, effectively minimized the overlap of high dose and high LET in OARs without sacrificing the physical dose distribution or its robustness in targets for prostate cancer patients treated with SSPT. DLVCRO employed a user-specified tool of DLVC, based on DLVH, to strategically redistribute high LET from OARs to targets, thereby potentially reducing the possible incidence rate of AEs in prostate cancer patients treated with SSPT. DLVCRO upgraded 2D DVH-based to 3D DLVH-based treatment planning to adjust dose/LET distributions simultaneously and robustly[8, 9] to reduce AEs in SSPT. DLVCRO is thus positioned as a front-runner for administering biologically optimized proton therapy.

# References

1. Eulitz J, Lutz B, Wohlfahrt P, Dutz A, Enghardt W, Karpowitz C, Krause M, Troost E, Lühr A. A Monte Carlo based radiation response modelling framework to assess variability of clinical RBE in proton therapy. Physics in Medicine & Biology. 2019;64(22):225020.
2. Mohan R, Grosshans D. Proton therapy–present and future. Advanced drug delivery reviews. 2017;109:26-44.
3. Smith AR. Vision: proton therapy. Medical physics. 2009;36(2):556-68.
4. Zhu XR, Li Y, Mackin D, Li H, Poenisch F, Lee AK, Mahajan A, Frank SJ, Gillin MT, Sahoo N. Towards effective and efficient patient-specific quality assurance for spot scanning proton therapy. Cancers. 2015;7(2):631-47.
5. Rwigema J-CM, Langendijk JA, van der Laan HP, Lukens JN, Swisher-McClure SD, Lin A. A model-based approach to predict short-term toxicity benefits with proton therapy for oropharyngeal cancer. International Journal of Radiation Oncology* Biology* Physics. 2019;104(3):553-62.
6. Blanchard P, Garden AS, Gunn GB, Rosenthal DI, Morrison WH, Hernandez M, Crutison J, Lee JJ, Ye R, Fuller CD. Intensity-modulated proton beam therapy (IMPT) versus intensity-modulated photon therapy (IMRT) for patients with oropharynx cancer–a case matched analysis. Radiotherapy and Oncology. 2016;120(1):48-55.
7. Blanchard P, Wong AJ, Gunn GB, Garden AS, Mohamed AS, Rosenthal DI, Crutison J, Wu R, Zhang X, Zhu XR. Toward a model-based patient selection strategy for proton therapy: external validation of photon-derived normal tissue complication probability models in a head and neck proton therapy cohort. Radiotherapy and Oncology. 2016;121(3):381-6.
8. An Y, Shan J, Patel SH, Wong W, Schild SE, Ding X, Bues M, Liu W. Robust intensity-modulated proton therapy to reduce high linear energy transfer in organs at risk. Med Phys. 2017;44(12):6138-47. Epub 2017/10/05. doi: 10.1002/mp.12610. PubMed PMID: 28976574; PMCID: PMC5734644.
9. Liu C, Patel SH, Shan J, Schild SE, Vargas CE, Wong WW, Ding X, Bues M, Liu W. Robust Optimization for Intensity Modulated Proton Therapy to Redistribute High Linear Energy Transfer from Nearby Critical Organs to Tumors in Head and Neck Cancer. International journal of radiation oncology, biology, physics. 2020;107(1):181-93. Epub 2020/01/29. doi: 10.1016/j.ijrobp.2020.01.013. PubMed PMID: 31987967.
10. Liu W, Zhang X, Li Y, Mohan R. Robust optimization in intensity-modulated proton therapy. Med Phys. 2012;39:1079-91.
11. An Y, Liang J, Schild SE, Bues M, Liu W. Robust treatment planning with conditional value at risk chance constraints in intensity‐modulated proton therapy. Medical physics. 2017;44(1):28-36.
12. Liu C, Bhangoo RS, Sio TT, Yu NY, Shan J, Chiang JS, Ding JX, Rule WG, Korte S, Lara P, Ding X, Bues M, Hu Y, DeWees T, Ashman JB, Liu W. Dosimetric comparison of distal esophageal carcinoma plans for patients treated with small-spot intensity-modulated proton versus volumetric-modulated arc therapies. J Appl Clin Med Phys. 2019;20(7):15-27. Epub 2019/05/22. doi: 10.1002/acm2.12623. PubMed PMID: 31112371; PMCID: PMC6612702.
13. Liu C, Sio TT, Deng W, Shan J, Daniels TB, Rule WG, Lara PR, Korte SM, Shen J, Ding X, Schild SE, Bues M, Liu W. Small-spot intensity-modulated proton therapy and volumetric-modulated arc therapies for patients with locally advanced non-small-cell lung cancer: A dosimetric comparative study. J Appl Clin Med Phys. 2018;19(6):140-8. Epub 2018/10/18. doi: 10.1002/acm2.12459. PubMed PMID: 30328674; PMCID: PMC6236833.
14. Liu CB, Schild SE, Chang JY, Liao ZX, Korte S, Shen JJ, Ding XN, Hu YL, Kang YX, Keole SR, Sio TT, Wong WW, Sahoo N, Bues M, Liu W. Impact of Spot Size and Spacing on the Quality of Robustly Optimized Intensity Modulated Proton Therapy Plans for Lung Cancer. International Journal of Radiation Oncology Biology

Physics. 2018;101(2):479-89. doi: 10.1016/j.ijrobp.2018.02.009. PubMed PMID: WOS:000432448900036.
15. Liu W, Mohan R, Park P, Liu Z, Li H, Li X, Li Y, Wu R, Sahoo N, Dong L, Zhu XR, Grosshans DR. Dosimetric benefits of robust treatment planning for intensity modulated proton therapy for base-of-skull cancers. Practical Radiation Oncology. 2014;4:384-91.
16. Liu W, Schild SE, Chang JY, Liao ZX, Chang YH, Wen ZF, Shen JJ, Stoker JB, Ding XN, Hu YL, Sahoo N, Herman MG, Vargas C, Keole S, Wong W, Bues M. Exploratory Study of 4D versus 3D Robust Optimization in Intensity Modulated Proton Therapy for Lung Cancer. International Journal of Radiation Oncology Biology Physics. 2016;95(1):523-33. doi: 10.1016/j.ijrobp.2015.11.002. PubMed PMID: WOS:000375419500072.
17. Fredriksson A, Forsgren A, Hårdemark B. Minimax optimization for handling range and setup uncertainties in proton therapy. Medical physics. 2011;38(3):1672-84.
18. Lomax A. Intensity modulated proton therapy and its sensitivity to treatment uncertainties 2: the potential effects of inter-fraction and inter-field motions. Physics in Medicine & Biology. 2008;53(4):1043.
19. Lomax A. Intensity modulated proton therapy and its sensitivity to treatment uncertainties 1: the potential effects of calculational uncertainties. Physics in Medicine & Biology. 2008;53(4):1027.
20. Pflugfelder D, Wilkens J, Oelfke U. Worst case optimization: a method to account for uncertainties in the optimization of intensity modulated proton therapy. Physics in Medicine & Biology. 2008;53(6):1689.
21. Unkelbach J, Alber M, Bangert M, Bokrantz R, Chan TCY, Deasy JO, Fredriksson A, Gorissen BL, van Herk M, Liu W, Mahmoudzadeh H, Nohadani O, Siebers JV, Witte M, Xu H. Robust radiotherapy planning. Phys Med Biol. 2018;63(22):22TR02. Epub 2018/11/13. doi: 10.1088/1361-6560/aae659. PubMed PMID: 30418942.
22. Paganetti H, Niemierko A, Ancukiewicz M, Gerweck LE, Goitein M, Loeffler JS, Suit HD. Relative biological effectiveness (RBE) values for proton beam therapy. International Journal of Radiation Oncology* Biology* Physics. 2002;53(2):407-21.
23. Unkelbach J, Botas P, Giantsoudi D, Gorissen BL, Paganetti H. Reoptimization of Intensity Modulated Proton Therapy Plans Based on Linear Energy Transfer. International journal of radiation oncology, biology, physics. 2016;96(5):1097-106. Epub 2016/11/22. doi: 10.1016/j.ijrobp.2016.08.038. PubMed PMID: 27869082; PMCID: PMC5133459.
24. Peeler CR, Mirkovic D, Titt U, Blanchard P, Gunther JR, Mahajan A, Mohan R, Grosshans DR. Clinical evidence of variable proton biological effectiveness in pediatric patients treated for ependymoma. Radiotherapy and Oncology. 2016;121(3):395-401.
25. Underwood TS, Grassberger C, Bass R, MacDonald SM, Meyersohn NM, Yeap BY, Jimenez RB, Paganetti H. Asymptomatic late-phase radiographic changes among chest-wall patients are associated with a proton RBE exceeding 1.1. International Journal of Radiation Oncology* Biology* Physics. 2018;101(4):809-19.
26. Ödén J, Toma‐Dasu I, Witt Nyström P, Traneus E, Dasu A. Spatial correlation of linear energy transfer and relative biological effectiveness with suspected treatment‐related toxicities following proton therapy for intracranial tumors. Medical physics. 2020;47(2):342-51.
27. Paganetti H. Relative biological effectiveness (RBE) values for proton beam therapy. Variations as a function of biological endpoint, dose, and linear energy transfer. Physics in Medicine & Biology. 2014;59(22):R419.
28. Cao W, Khabazian A, Yepes PP, Lim G, Poenisch F, Grosshans DR, Mohan R. Linear energy transfer incorporated intensity modulated proton therapy optimization. Physics in Medicine & Biology. 2017;63(1):015013.
29. Guan F, Bronk L, Titt U, Lin SH, Mirkovic D, Kerr MD, Zhu XR, Dinh J, Sobieski M, Stephan C. Spatial mapping of the biologic effectiveness of scanned particle beams: towards biologically optimized particle therapy. Scientific reports. 2015;5(1):9850.

30. Patel D, Bronk L, Guan F, Peeler CR, Brons S, Dokic I, Abdollahi A, Rittmüller C, Jäkel O, Grosshans D. Optimization of Monte Carlo particle transport parameters and validation of a novel high throughput experimental setup to measure the biological effects of particle beams. Medical physics. 2017;44(11):6061-73.
31. Chaudhary P, Marshall TI, Perozziello FM, Manti L, Currell FJ, Hanton F, McMahon SJ, Kavanagh JN, Cirrone GAP, Romano F. Relative biological effectiveness variation along monoenergetic and modulated Bragg peaks of a 62-MeV therapeutic proton beam: a preclinical assessment. International Journal of Radiation Oncology* Biology* Physics. 2014;90(1):27-35.
32. Gentile MS, Yeap BY, Paganetti H, Goebel CP, Gaudet DE, Gallotto SL, Weyman EA, Morgan ML, MacDonald SM, Giantsoudi D. Brainstem injury in pediatric patients with posterior fossa tumors treated with proton beam therapy and associated dosimetric factors. International Journal of Radiation Oncology* Biology* Physics. 2018;100(3):719-29.
33. Bolsi A, Placidi L, Pica A, Ahlhelm FJ, Walser M, Lomax AJ, Weber DC. Pencil beam scanning proton therapy for the treatment of craniopharyngioma complicated with radiation-induced cerebral vasculopathies: a dosimetric and linear energy transfer (LET) evaluation. Radiotherapy and Oncology. 2020;149:197-204.
34. Fager M, Toma-Dasu I, Kirk M, Dolney D, Diffenderfer ES, Vapiwala N, Carabe A. Linear energy transfer painting with proton therapy: a means of reducing radiation doses with equivalent clinical effectiveness. International Journal of Radiation Oncology* Biology* Physics. 2015;91(5):1057-64.
35. Carabe A, Espana S, Grassberger C, Paganetti H. Clinical consequences of relative biological effectiveness variations in proton radiotherapy of the prostate, brain and liver. Physics in Medicine & Biology. 2013;58(7):2103.
36. Beltran C, Tseung HWC, Augustine KE, Bues M, Mundy DW, Walsh TJ, Herman MG, Laack NN. Clinical implementation of a proton dose verification system utilizing a GPU accelerated Monte Carlo engine. International Journal of Particle Therapy. 2016;3(2):312-9.
37. Rørvik E, Thörnqvist S, Stokkevåg CH, Dahle TJ, Fjæra LF, Ytre‐Hauge KS. A phenomenological biological dose model for proton therapy based on linear energy transfer spectra. Medical physics. 2017;44(6):2586-94.
38. Hawkins RB. A microdosimetric-kinetic model for the effect of non-Poisson distribution of lethal lesions on the variation of RBE with LET. Radiation research. 2003;160(1):61-9.
39. Schulz-Ertner D, Karger CP, Feuerhake A, Nikoghosyan A, Combs SE, Jäkel O, Edler L, Scholz M, Debus J. Effectiveness of carbon ion radiotherapy in the treatment of skull-base chordomas. International Journal of Radiation Oncology* Biology* Physics. 2007;68(2):449-57.
40. Elsässer T, Scholz M. Cluster effects within the local effect model. Radiation research. 2007;167(3):319-29.
41. Elsässer T, Krämer M, Scholz M. Accuracy of the local effect model for the prediction of biologic effects of carbon ion beams in vitro and in vivo. International Journal of Radiation Oncology* Biology* Physics. 2008;71(3):866-72.
42. Elsässer T, Weyrather WK, Friedrich T, Durante M, Iancu G, Krämer M, Kragl G, Brons S, Winter M, Weber K-J. Quantification of the relative biological effectiveness for ion beam radiotherapy: direct experimental comparison of proton and carbon ion beams and a novel approach for treatment planning. International Journal of Radiation Oncology* Biology* Physics. 2010;78(4):1177-83.
43. Carlson DJ, Stewart RD, Semenenko VA, Sandison GA. Combined use of Monte Carlo DNA damage simulations and deterministic repair models to examine putative mechanisms of cell killing. Radiation research. 2008;169(4):447-59.
44. Wu Q, Mohan R. Algorithms and functionality of an intensity modulated radiotherapy optimization system. Med Phys. 2000;27(4):701-11. Epub 2000/05/08. doi: 10.1118/1.598932. PubMed PMID: 10798692.

45. Giantsoudi D, Grassberger C, Craft D, Niemierko A, Trofimov A, Paganetti H. Linear energy transfer-guided optimization in intensity modulated proton therapy: feasibility study and clinical potential. International Journal of Radiation Oncology* Biology* Physics. 2013;87(1):216-22.
46. Bassler N, Jäkel O, Søndergaard CS, Petersen JB. Dose-and LET-painting with particle therapy. Acta oncologica. 2010;49(7):1170-6.
47. Bassler N, Toftegaard J, Lühr A, Sørensen BS, Scifoni E, Krämer M, Jäkel O, Mortensen LS, Overgaard J, Petersen JB. LET-painting increases tumour control probability in hypoxic tumours. Acta oncologica. 2014;53(1):25-32.
48. Inaniwa T, Kanematsu N, Noda K, Kamada T. Treatment planning of intensity modulated composite particle therapy with dose and linear energy transfer optimization. Physics in Medicine & Biology. 2017;62(12):5180.
49. Traneus E, Ödén J. Introducing proton track-end objectives in intensity modulated proton therapy optimization to reduce linear energy transfer and relative biological effectiveness in critical structures. International Journal of Radiation Oncology* Biology* Physics. 2019;103(3):747-57.
50. Liu R, Charyyev S, Wahl N, Liu W, Kang M, Zhou J, Yang X, Baltazar F, Palkowitsch M, Higgins K. An Integrated Biological Optimization framework for proton SBRT FLASH treatment planning allows dose, dose rate, and LET optimization using patient-specific ridge filters. arXiv preprint arXiv:220708016. 2022.
51. Harrison N, Kang M, Liu R, Charyyev S, Wahl N, Liu W, Zhou J, Higgins KA, Simone II CB, Bradley JD. A novel inverse algorithm to solve IPO-IMPT of proton FLASH therapy with sparse filters. International Journal of Radiation Oncology* Biology* Physics. 2023.
52. Deng W, Yang Y, Liu C, Bues M, Mohan R, Wong WW, Foote RH, Patel SH, Liu W. A Critical Review of LET-Based Intensity-Modulated Proton Therapy Plan Evaluation and Optimization for Head and Neck Cancer Management. Int J Part Ther. 2021;8(1):36-49. Epub 2021/07/22. doi: 10.14338/IJPT-20-00049.1. PubMed PMID: 34285934; PMCID: PMC8270082.
53. Rørvik E, Fjæra LF, Dahle TJ, Dale JE, Engeseth GM, Stokkevåg CH, Thörnqvist S, Ytre-Hauge KS. Exploration and application of phenomenological RBE models for proton therapy. Physics in Medicine & Biology. 2018;63(18):185013.
54. McIntyre M, Wilson P, Gorayski P, Bezak E. A Systematic Review of LET-Guided Treatment Plan Optimisation in Proton Therapy: Identifying the Current State and Future Needs. Cancers (Basel). 2023;15(17). Epub 20230825. doi: 10.3390/cancers15174268. PubMed PMID: 37686544; PMCID: PMC10486456.
55. Yang Y, Vargas CE, Bhangoo RS, Wong WW, Schild SE, Daniels TB, Keole SR, Rwigema JM, Glass JL, Shen J, DeWees TA, Liu T, Bues M, Fatyga M, Liu W. Exploratory Investigation of Dose-Linear Energy Transfer (LET) Volume Histogram (DLVH) for Adverse Events Study in Intensity Modulated Proton Therapy (IMPT). International journal of radiation oncology, biology, physics. 2021;110(4):1189-99. Epub 2021/02/24. doi: 10.1016/j.ijrobp.2021.02.024. PubMed PMID: 33621660.
56. Yang Y, Patel SH, Bridhikitti J, Wong WW, Halyard MY, McGee LA, Rwigema JM, Schild SE, Vora SA, Liu T, Bues M, Fatyga M, Foote RL, Liu W. Exploratory study of seed spots analysis to characterize dose and linear-energy-transfer effect in adverse event initialization of pencil-beam-scanning proton therapy. Med Phys. 2022;49(9):6237-52. Epub 2022/07/13. doi: 10.1002/mp.15859. PubMed PMID: 35820062.
57. Yang Y, Gergelis KR, Shen J, Afzal A, Mullikin TC, Gao RW, Aziz K, Shumway DA, Corbin KS, Liu W. Study of linear energy transfer effect on rib fracture in breast patients receiving pencil-beamscanning proton therapy. ArXiv. 2023.
58. Aarup-Kristensen S, Hansen CR, Forner L, Brink C, Eriksen JG, Johansen J. Osteoradionecrosis of the mandible after radiotherapy for head and neck cancer: risk factors and dose-volume correlations. Acta oncologica (Stockholm, Sweden). 2019;58(10):1373-7. Epub 2019/08/01. doi: 10.1080/0284186X.2019.1643037. PubMed PMID: 31364903.

59. Kuhnt T, Stang A, Wienke A, Vordermark D, Schweyen R, Hey J. Potential risk factors for jaw osteoradionecrosis after radiotherapy for head and neck cancer. Radiation oncology (London, England). 2016;11:101-. doi: 10.1186/s13014-016-0679-6. PubMed PMID: 27473433.
60. Gergelis K, Yang Y, Afzal A, Mullikin T, Vargas C, McGee L, Ahmed S, Park S, Shumway D, Corbin K. Study of Linear Energy Transfer Effect on Rib Fracture in Breast Patients Receiving Pencil-Beam-Scanning Proton Therapy. International Journal of Radiation Oncology* Biology* Physics. 2022;114(3):e29.
61. Marteinsdottir M, Wang C-C, McNamara AL, Depauw N, Shin J, Paganetti H. The impact of variable RBE in proton therapy for left-sided breast cancer when estimating normal tissue complications in the heart and lung. Physics in Medicine and Biology. 2020.
62. Frankart AJ, Frankart MJ, Cervenka B, Tang AL, Krishnan DG, Takiar V. Osteoradionecrosis: Exposing the evidence not the bone. International Journal of Radiation Oncology* Biology* Physics. 2021;109(5):1206-18.
63. Feng H, Shan J, Anderson JD, Wong WW, Schild SE, Foote RL, Patrick CL, Tinnon KB, Fatyga M, Bues M, Patel SH, Liu W. Per-voxel constraints to minimize hot spots in linear energy transfer-guided robust optimization for base of skull head and neck cancer patients in IMPT. Med Phys. 2022;49(1):632-47. Epub 2021/11/30. doi: 10.1002/mp.15384. PubMed PMID: 34843119.
64. Drzymala R, Mohan R, Brewster L, Chu J, Goitein M, Harms W, Urie M. Dose-volume histograms. International Journal of Radiation Oncology* Biology* Physics. 1991;21(1):71-8.
65. Schneider U, Pedroni E, Lomax A. The calibration of CT Hounsfield units for radiotherapy treatment planning. Physics in Medicine & Biology. 1996;41(1):111.
66. Schaffner B, Pedroni E. The precision of proton range calculations in proton radiotherapy treatment planning: experimental verification of the relation between CT-HU and proton stopping power. Physics in Medicine & Biology. 1998;43(6):1579.
67. Unkelbach J, Bortfeld T, Martin BC, Soukup M. Reducing the sensitivity of IMPT treatment plans to setup errors and range uncertainties via probabilistic treatment planning. Medical physics. 2009;36(1):149-63.
68. Unkelbach J, Chan TC, Bortfeld T. Accounting for range uncertainties in the optimization of intensity modulated proton therapy. Physics in Medicine & Biology. 2007;52(10):2755.
69. Wilkens J, Oelfke U. A phenomenological model for the relative biological effectiveness in therapeutic proton beams. Physics in Medicine & Biology. 2004;49(13):2811.
70. Guan F, Peeler C, Bronk L, Geng C, Taleei R, Randeniya S, Ge S, Mirkovic D, Grosshans D, Mohan R. Analysis of the track‐ and dose‐averaged LET and LET spectra in proton therapy using the geant4 Monte Carlo code. Medical physics. 2015;42(11):6234-47.
71. Bai X, Lim G, Grosshans D, Mohan R, Cao W. Robust optimization to reduce the impact of biological effect variation from physical uncertainties in intensity-modulated proton therapy. Phys Med Biol. 2019;64(2):025004. Epub 2018/12/14. doi: 10.1088/1361-6560/aaf5e9. PubMed PMID: 30523932.
72. Liu W, Zhang X, Li Y, Mohan R. Robust optimization of intensity modulated proton therapy. Med Phys. 2012;39(2):1079-91. Epub 2012/02/11. doi: 10.1118/1.3679340. PubMed PMID: 22320818; PMCID: PMC3281975.
73. Yang Y, Muller OM, Shiraishi S, Harper M, Amundson AC, Wong WW, McGee LA, Rwigema JM, Schild SE, Bues M, Fatyga M, Anderson JD, Patel SH, Foote RL, Liu W. Empirical Relative Biological Effectiveness (RBE) for Mandible Osteoradionecrosis (ORN) in Head and Neck Cancer Patients Treated With Pencil-Beam-Scanning Proton Therapy (PBSPT): A Retrospective, Case-Matched Cohort Study. Front Oncol. 2022;12:843175. Epub 2022/03/22. doi: 10.3389/fonc.2022.843175. PubMed PMID: 35311159; PMCID: PMC8928456.
74. Gu W, Ruan D, Zou W, Dong L, Sheng K. Linear energy transfer weighted beam orientation optimization for intensity‐modulated proton therapy. Medical physics. 2021;48(1):57-70.

75. An Y, Liang J, Schild SE, Bues M, Liu W. Robust treatment planning with conditional value at risk chance constraints in intensity-modulated proton therapy. Med Phys. 2017;44(1):28-36. Epub 2017/01/04. doi: 10.1002/mp.12001. PubMed PMID: 28044325; PMCID: PMC5388360.
76. Traneus E, Oden J. Introducing Proton Track-End Objectives in Intensity Modulated Proton Therapy Optimization to Reduce Linear Energy Transfer and Relative Biological Effectiveness in Critical Structures. International journal of radiation oncology, biology, physics. 2019;103(3):747-57. Epub 2018/11/06. doi: 10.1016/j.ijrobp.2018.10.031. PubMed PMID: 30395906.
77. An Y, Shan J, Patel SH, Wong W, Schild SE, Ding X, Bues M, Liu W. Robust intensity‐modulated proton therapy to reduce high linear energy transfer in organs at risk. Medical physics. 2017;44(12):6138-47.
78. Unkelbach J, Botas P, Giantsoudi D, Gorissen BL, Paganetti H. Reoptimization of intensity modulated proton therapy plans based on linear energy transfer. International Journal of Radiation Oncology* Biology* Physics. 2016;96(5):1097-106.
79. Fager M, Toma-Dasu I, Kirk M, Dolney D, Diffenderfer ES, Vapiwala N, Carabe A. Linear energy transfer painting with proton therapy: A means of reducing radiation doses with equivalent clinical effectiveness. International Journal of Radiation Oncology• Biology• Physics. 2015;91(5):1057-64.
80. Giantsoudi D, Grassberger C, Craft D, Niemierko A, Trofimov A, Paganetti H. Linear energy transfer-guided optimization in intensity modulated proton therapy: feasibility study and clinical potential. International Journal of Radiation Oncology• Biology• Physics. 2013;87(1):216-22.
81. Shan J, An Y, Bues M, Schild SE, Liu W. Robust optimization in IMPT using quadratic objective functions to account for the minimum MU constraint. Med Phys. 2018;45(1):460-9. Epub 2017/11/18. doi: 10.1002/mp.12677. PubMed PMID: 29148570; PMCID: PMC5774242.
82. Shan J, Sio TT, Liu C, Schild SE, Bues M, Liu W. A novel and individualized robust optimization method using normalized dose interval volume constraints (NDIVC) for intensity-modulated proton radiotherapy. Med Phys. 2019;46(1):382-93. Epub 2018/11/06. doi: 10.1002/mp.13276. PubMed PMID: 30387870.
83. Liu W, Li Y, Li X, Cao W, Zhang X. Influence of robust optimization in intensity-modulated proton therapy with different dose delivery techniques. Med Phys. 2012;39(6):3089-101. Epub 2012/07/05. doi: 10.1118/1.4711909. PubMed PMID: 22755694; PMCID: PMC3360691.
84. Liu W. Robustness Quantification and Worst-Case Robust Optimization in Intensity-Modulated Proton Therapy.  Particle Radiotherapy2016. p. 139-55.
85. Liu W, Liao Z, Schild SE, Liu Z, Li H, Li Y, Park PC, Li X, Stoker J, Shen J, Keole S, Anand A, Fatyga M, Dong L, Sahoo N, Vora S, Wong W, Zhu XR, Bues M, Mohan R. Impact of respiratory motion on worst-case scenario optimized intensity modulated proton therapy for lung cancers. Practical Radiation Oncology. 2015;5(2):e77-86. doi: 10.1016/j.prro.2014.08.002. PubMed PMID: MEDLINE:25413400.
86. Ding Y, Feng H, Yang Y, Holmes J, Liu Z, Liu D, Wong WW, Yu NY, Sio TT, Schild SE, Li B, Liu W. Deep-learning based fast and accurate 3D CT deformable image registration in lung cancer. Medical Physics.n/a(n/a). doi: https://doi.org/10.1002/mp.16548.
87. Shan J, Sio TT, Liu C, Schild SE, Bues M, Liu W. A novel and individualized robust optimization method using normalized dose interval volume constraints (NDIVC) for intensity-modulated proton radiotherapy. Med Phys. 2018. Epub 2018/11/06. doi: 10.1002/mp.13276. PubMed PMID: 30387870.
88. Moyers MF, Miller DW, Bush DA, Slater JD. Methodologies and tools for proton beam design for lung tumors. International journal of radiation oncology, biology, physics. 2001;49(5):1429-38. Epub 2001/04/05. doi: 10.1016/s0360-3016(00)01555-8. PubMed PMID: 11286851.
89. Yang M, Zhu XR, Park PC, Titt U, Mohan R, Virshup G, Clayton JE, Dong L. Comprehensive analysis of proton range uncertainties related to patient stopping-power-ratio estimation using the stoichiometric calibration. Phys Med Biol. 2012;57(13):4095-115. Epub 2012/06/09. doi: 10.1088/0031-9155/57/13/4095. PubMed PMID: 22678123; PMCID: PMC3396587.

90. Liu W, Frank SJ, Li X, Li Y, Park PC, Dong L, Ronald Zhu X, Mohan R. Effectiveness of robust optimization in intensity-modulated proton therapy planning for head and neck cancers. Med Phys. 2013;40(5):051711. Epub 2013/05/03. doi: 10.1118/1.4801899. PubMed PMID: 23635259; PMCID: PMC3651255.
91. Lomax AJ. Intensity modulated proton therapy: the potential and the challenge [Habilitation Thesis]. Zurich: ETH; 2004.
92. Younkin JE, Morales DH, Shen J, Shan J, Bues M, Lentz JM, Schild SE, Stoker JB, Ding X, Liu W. Clinical Validation of a Ray-Casting Analytical Dose Engine for Spot Scanning Proton Delivery Systems. Technology in Cancer Research & Treatment. 2019;18. doi: 10.1177/1533033819887182.
93. Deng W, Ding X, Younkin JE, Shen J, Bues M, Schild SE, Patel SH, Liu W. Hybrid 3D analytical linear energy transfer calculation algorithm based on precalculated data from Monte Carlo simulations. Med Phys. 2020;47(2):745-52. Epub 2019/11/24. doi: 10.1002/mp.13934. PubMed PMID: 31758864.
94. Holmes J, Shen J, Shan J, Patrick CL, Wong WW, Foote RL, Patel SH, Bues M, Liu W. Evaluation and second check of a commercial Monte Carlo dose engine for small‐field apertures in pencil beam scanning proton therapy. Medical physics. 2022;49(5):3497-506.
95. Shan J, Feng H, Morales DH, Patel SH, Wong WW, Fatyga M, Bues M, Schild SE, Foote RL, Liu W. Virtual particle Monte Carlo: A new concept to avoid simulating secondary particles in proton therapy dose calculation. Medical physics. 2022;49(10):6666-83.
96. Hongying Feng WL. Modelling small block aperture in an in-house developed GPU-accelerated Monte Carlo-based dose engine for pencil beam scanning proton therapy. Arxiv. 2023.
97. Liu DC, Nocedal J. On the limited memory BFGS method for large scale optimization. Mathematical programming. 1989;45(1):503-28.
98. Trofimov A, Unkelbach J, DeLaney TF, Bortfeld T. Visualization of a variety of possible dosimetric outcomes in radiation therapy using dose-volume histogram bands. Practical radiation oncology. 2012;2(3):164-71.
99. Casiraghi M, Albertini F, Lomax AJ. Advantages and limitations of the 'worst case scenario' approach in IMPT treatment planning. Physics in Medicine & Biology. 2013;58(5):1323.
100. Wilkens JJ, Oelfke U. Optimization of radiobiological effects in intensity modulated proton therapy. Medical physics. 2005;32(2):455-65.
101. Frese MC, Wilkens JJ, Huber PE, Jensen AD, Oelfke U, Taheri-Kadkhoda Z. Application of constant vs. variable relative biological effectiveness in treatment planning of intensity-modulated proton therapy. International Journal of Radiation Oncology* Biology* Physics. 2011;79(1):80-8.
102. Grassberger C, Trofimov A, Lomax A, Paganetti H. Variations in linear energy transfer within clinical proton therapy fields and the potential for biological treatment planning. International Journal of Radiation Oncology* Biology* Physics. 2011;80(5):1559-66.
103. Grassberger C, Paganetti H. Elevated LET components in clinical proton beams. Physics in Medicine & Biology. 2011;56(20):6677.
104. Wilkens JJ, Oelfke U. Three-dimensional LET calculations for treatment planning of proton therapy. Zeitschrift für medizinische Physik. 2004;14(1):41-6.